# Critical Field Determination in Organic Antiferromagnet F4BImNN.

Cris Adriano[1], Rafael S. Freitas*[2], Armando Paduan-Filho[2], Pascoal G. Pagliuso[1], Nei Fernandes Oliveira Jr.‡[2], Paul M. Lahti*‡[3]

[1]Instituto de Física Gleb Wataghin, Universidade Estadual de Campinas, Campinas, SP, Brazil
[2]Instituto de Física, Universidade de São Paulo, São Paulo, SP, Brazil
[3]Department of Chemistry, University of Massachusetts, Amherst, MA 01003 USA

**Abstract:** A combination of magnetic heat capacity measurements versus temperature under multiple fixed external fields using a polycrystalline sample, plus easy and hard axis aligned magnetization versus field measurements using a single crystal of the radical 2-(4,5,6,7-tetrafluorobenzimidazol-2-yl)-4,4,5,5-tetramethyl-4,5-dihydro-1$H$-imidazole-3-oxide-1-oxyl) at multiple fixed temperatures ≤0.75 K, confirmed a zero-field Néel temperature of $T_N(0)$ = 0.73 K, with extrapolated 0 K critical field parameters of $H_c(\|)$ = 170 Oe and $H_c(\perp)$ = 306 Oe. Assuming the previously determined 1-D intrachain ferromagnetic exchange interaction of $J/k$ = +22 K between radicals (Hamiltonian H = –2J $\Sigma S_i \cdot S_j$), the interchain exchange in antiferromagnetic with $2zJ_{inter}$ = –0.09 K.

In 2008 some of us (with others) reported[1] a magnetostructural study of 2-(4,5,6,7-tetrafluorobenzimidazol-2-yl)-4,4,5,5-tetramethyl-4,5-dihydro-$1H$-imidazole-3-oxide-1-oxyl (F4BImNN). During subsequent experiments with F4BImNN, we came to consider its apparent critical field, $H_c$ = 1.81±0.07 kOe, to be surprisingly high. We therefore decided to re-examine the heat capacity behavior of F4BImNN in an external magnetic field near its Néel temperature of $T_N$ = 0.7 K. In addition to doing new heat capacity experiments, we also carried out magnetic measurements to allow comparison of *both* magnetic and calorimetric results for F4BImNN from the same sample material.

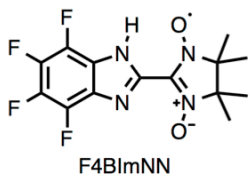

F4BImNN

Ac magnetic susceptibility measurements were carried out using a custom-built apparatus that has been described[2] elsewhere. The measurements were made at several fixed temperatures near and below the zero-field Néel temperature of $T_N(0)$ = 0.7 K, in varying external magnetic field with the crystal in both easy axis and hard axis alignments. These were done using a single prism crystal (3×1×0.3 mm) of F4BImNN immobilized with grease in a PVC sample holder. The data were obtained using a 155 Hz ac-field modulation frequency with a 5 Oe modulation amplitude.

Plots of susceptibility versus field plot maxima for these experiments give the spin-flip transition field for each temperature at or below $T_N(0)$, as exemplified in Figure 1. In both easy and hard axis orientations, the spin-flip transition field increased at lower temperatures, as will be described in more detail below.

The heat capacity $C_P$ of F4BImNN from the same sample batch was studied over a temperature range of 0.1–2.0 K at various fixed external dc-magnetic fields, using a sample of multiple crystals with easy axes aligned parallel to the applied field direction in a Quantum Design Physical Property Measurement System (PPMS) with a $^3$He dilution refrigerator option and standard semiadiabatic heat pulse technique.

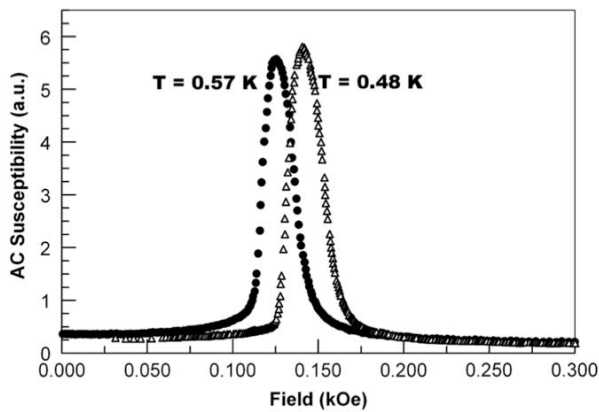

**Figure 1**. Example ac magnetic susceptibility (arbitrary units) vs. external magnetic field for a single crystal of F4BImNN (easy axis alignment); 155 Hz ac-field modulation frequency with 5 Oe modulation amplitude.

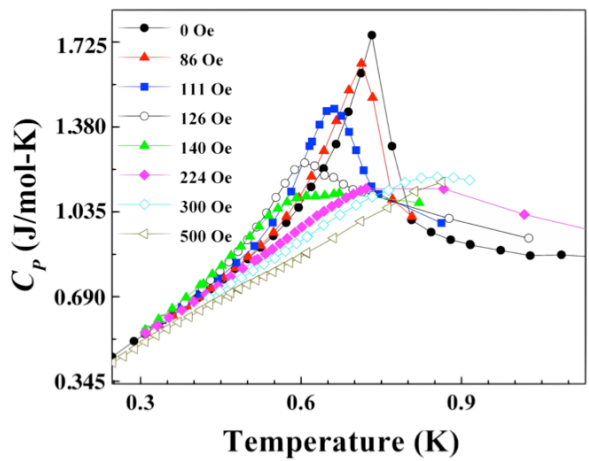

**Figure 2**. (a) Heat capacity $C_P$ versus temperature data for a polycrystalline sample of F4BImNN measured at several fixed external magnetic fields.

The calorimetric results are shown in Figure 2. The shapes of the re-examined $C_P$ versus $T$ curves at the various external fields are very consistent with the curves seen in the original[1] study, as are the changes in shape as external field increases. In particular, the lower field data show clear cusp maxima that indicate long range magnetic ordering. At higher fields, the cusp maxima shift to lower temperature and get broader, until only a broad baseline feature remains, in accord with expectations[3] for a low dimensional, 1-D chain system forming an antiferromagnetic ordered phase under these conditions. The zero-field cusp for $T_N(0)$ occurs at 0.73 K, in excellent agreement with the cusp position in the original paper used to determine $T_N(0)$.

Despite the qualitative shape similarities, the shape changes in the new $C_P$ versus $T$ curves occur at significantly lower external fields than seen[1] in the original report. Since the original report, service work on the instrumentation used to carry out the original magneto-calorimetry study showed a problem with the external magnetic field controller, which erroneously gave field value determinations that were too high. This problem is consistent with the discrepancy seen between the original $C_p$ vs $T$ results in nonzero external fields, and the present results in Figure 2. None of the zero-field measurements for the original report should have been affected, and indeed the new zero-field results correspond very well with the original zero-field results.

As a result of the new studies, for the first time both magnetic and calorimetric determinations of Néel temperature variation with external magnetic field are available for the same sample batch of F4BImNN. The combination of results is stronger than either set of data alone. In particular, because the strong 1-D FM exchange alignment axis in F4BImNN lies parallel to the crystal easy axis, it was straightforward to determine the Néel temperatures at multiple external fields. By comparison, calorimetry data at higher fields in a low-dimensionality material looses the clear cusp maxima from which the Néel temperature is readily determined. This is seen in the Figure 2 higher external field data.

So, the data from the new ac-magnetic data exemplified in Figure 1 were combined with the cusp maximum determinations in Figure 2 to construct the magnetic phase diagram in Figure 3 showing Néel temperatures versus the spin-flip field $H$. The combined parallel orientation data (lower curve), and the magnetic-only perpendicular orientation data were both fitted to equation 1,

$$T_N(H) = T_N(0)[1 - (H/H_c)^\alpha]^\xi \quad (1)$$

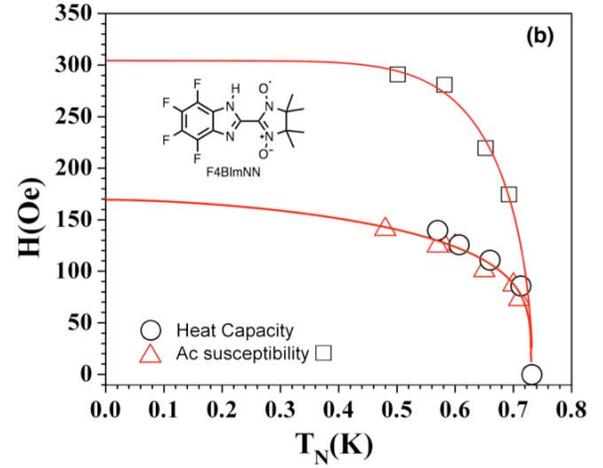

**Figure 3.** Magnetic phase diagram of antiferromagnetic transition fields versus Néel ordering temperatures determined from heat capacity cusp maxima in Figure 2 (○), and from magnetic susceptibility maxima (e.g., Figure 1) for an F4BImNN single crystal oriented with the easy axis parallel (△) and perpendicular (□) to the field direction. The solid lines are fits to equation (1) in the main text.

where $T_N(0)$ is the zero field Néel temperature, $H_c$ the critical field extrapolated to 0 K, with fitting coefficients $\alpha$ and $\xi$. This function is expected to give near-plateau behavior[4] of $H_c(T_N)$ at low $T_N$. For the parallel orientation data, the following parameters gave the fitted line that is shown in the lower curve of Figure 3: $H_c(\|) = 170$ Oe, $T_N(0) = 0.73$ K, $\alpha = 4.087$ and $\xi = 0.608$. The fitted $T_N(0)$ is very close to the observed zero-field $T_N(0)$, and the fitted $H_c$ is in good accord with the range of external field at which the ordering cusp is suppressed in the $C_P$ versus $T$ data of Figure 1. The perpendicular orientation data fit was done using the fixed experimental value $T_N(0) = 0.73$ K, and gave $H_c(\perp) = 306$ Oe, $\alpha = 1.836$ and $\xi = 0.135$.

The results for Figures 2-3 enable important refinement of the original critical field result for F4BImNN. Equation (2) relates[5] the critical field $H_c$ to the inter-chain antiferromagnetic interaction $2z|J_{inter}|$ between 1-D FM exchange-coupled spin chains in F4BImNN, where the spin quantum number $S = 1/2$, and the Landé factor $g = 2.0066$ from the previous[1] work. Using the value of $H_c(\|) = 170$ Oe obtained from this work, $2zJ_{inter} = -0.09$ K, with a negative value assigned due to the antiferromagnetic phase formation in F4BImNN.

$$2z|J_{inter}| = g\mu_B \cdot (H_c / S) \quad (2)$$

In summary, these follow-up results provide new and additional heat capacity and magnetic measurements yielding F4BImNN critical field determination: $H_c(\|) = 170$ Oe, $H_c(\perp) = 306$ Oe. The $H_c(\|)$ estimate provides a corrected value of $2zJ_{inter} = -0.09$ K for the interchain interaction between the strongly ferromagnetically coupled F4BImNN chains ($J/k = +22$ K with Hamiltonian $H = -2J \sum S_i \cdot S_j$)[1] whose nature is described in the original paper[1].

Given the limited number of such determinations for pure organic magnetic materials, we hope that these new, cross-checked results will prove useful for comparison to other low-dimensional organic molecular magnetic systems.

## AUTHOR INFORMATION

**Corresponding Author**

RSF -- freitas@if.usp.br
PML -- lahti@chem.umass.edu [‡]Emeritus, retired.

## ACKNOWLEDGMENT

This work was supported by FAPESP in Brazil (AP-F, RSF, NFO) grant : 2015/ 16191-5. We thank Prof. J. Mague of Tulane University for face-indexing a crystal of F4BImNN to identify the crystal easy axis as being parallel to its 1-D FM chain propagation axis. PML thanks APF for critical magnetic measurements on F4BImNN, and NFO for his determined pursuit of a self-consistent reconciliation of both magnetic and calorimetric measurements for F4BImNN.